\documentclass[conference,romanappendices]{IEEEtran}
\usepackage{amsmath}
\usepackage{graphics}
\usepackage{subfigure}
\usepackage{stfloats}
\usepackage{cite}
\usepackage{nomencl}
\usepackage{amssymb}
\usepackage{rotating}
\usepackage{psfrag}
\usepackage{color}
\usepackage{geometry}
\newtheorem{theorem}{\textbf{Theorem}}
\newtheorem{corollary}{\textbf{Corollary}}

\newtheorem{remark}{Remark}
\newcommand{\mc}{\mathcal} 

\begin{document}
\title{On Secure Transmission over Parallel Relay Eavesdropper Channel}
\author{\IEEEauthorblockN{Zohaib Hassan Awan, Abdellatif Zaidi and Luc Vandendorpe}
\IEEEauthorblockN{ ICTEAM institute, Universit\'{e} Catholique de Louvain, Louvain-la-Neuve 1348, Belgium.\\
E-mail: \{zohaib.awan, abdellatif.zaidi, luc.vandendorpe\}@uclouvain.be}%
}
\maketitle

\begin{abstract}
We study a four terminal parallel relay-eavesdropper channel which consists of multiple independent relay-eavesdropper channels as subchannels. For the discrete memoryless case, we establish inner and outer bounds on the rate-equivocation region. For each subchannel, secure transmission is obtained through one of the two coding schemes at the relay: decoding-and-forwarding the source message or confusing the eavesdropper through noise injection. The inner bound allows relay mode selection. For the Gaussian model we establish lower and upper bounds on the perfect secrecy rate. We show that the bounds meet in some special cases, including when the relay does not hear the source. We illustrate the analytical results through some numerical examples.
\end{abstract}
%

\section{Introduction}
The relay channel has been analyzed in \cite{11},\cite{12} (and references therein), but the focus was on how to increase achievable rate and reliability. The idea of cooperation between users in the context of security was introduced in \cite{10} (and references therein). The premise is that when the main channel is more noisy than the channel to the eavesdropper, cooperation between users is utilized to obtain a positive secrecy capacity. Secrecy is achieved by using the relay as a trusted node which facilitates the information decoding at the destination while confusing the eavesdropper.

\begin{figure}[ht]
\centering
\includegraphics[width=\linewidth]{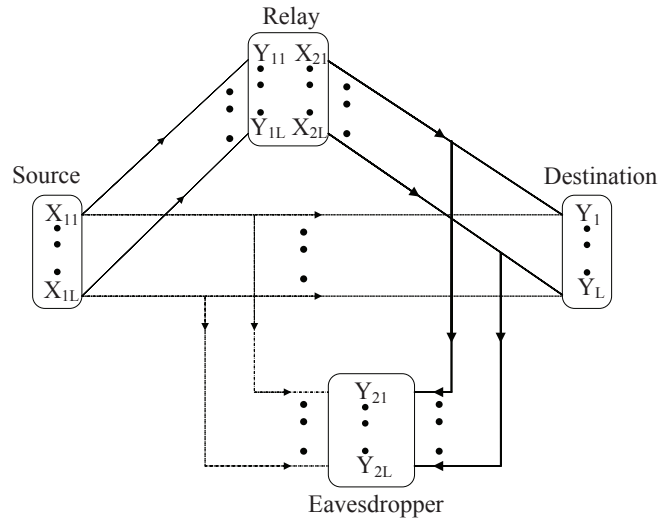}\\
 \caption{Parallel relay-eavesdropper channel.}
 \label{fig1}
\end{figure}

In this paper, we study a parallel relay-eavesdropper channel. A parallel relay-eavesdropper channel consists of a generalization of the setup in \cite{10} to the case in which each of the source-to-relay (S-R), source-to-destination (S-D), source-to-eavesdropper (S-E), relay-to-destination (R-D) and relay-to-eavesdropper (R-E) link is composed of several independent parallel channels as subchannels. The model is depicted in Figure \ref{fig1}. For this model, we establish outer and inner bounds on the rate-equivocation region for the discrete memoryless case. The inner bound is obtained with a coding scheme in which, for each subchannel, the relay operates in decode-and-forward (DF) or noise forwarding (NF) mode. The outer bound does not follow directly from the single-letter outer bound for the relay-eavesdropper channel developed in \cite[Theorem 1]{10} and so, a converse is needed. This converse includes a redefinition of the involved auxiliary random variables, a technique much similar to the one used before in the context of secure transmission over broadcast channels \cite{15}. We also show that the bounds on the equivocation rate coincide in the case in which all the subchannels are degraded, thus characterizing the secrecy capacity.

For the Gaussian model, we focus on the perfect secrecy case. We establish lower and upper bounds on the perfect secrecy rate. We note that establishing a computable upper bounds on the secrecy rate for the Gaussian model is non-trivial. In part, this is because converse techniques that are obtained directly from the analysis in the discrete case involves auxiliary random variables, the optimal choice of which is difficult to obtain. We develop a new upper bound on the secrecy rate for the parallel Gaussian relay-eavesdropper channel. Our converse proof uses elements from converse techniques developed in \cite{19},\cite{21} in the context of multi-antennas wiretap channel; and in a sense, can be viewed as a partial extension of these results to the case of the studied model. The established upper bound on the secrecy rate shows some degree of separability for different parallel subchannels. It is especially useful when the multiple access part of the channel is the bottleneck.

We also study a special Gaussian case in which the relay does not hear to the source, for example due to very noisy source-to-relay links. In this case, we show that noise-forwarding on all links achieves the secrecy capacity. The converse proof follows from the general converse established for the general Gaussian case, and a new genie-aided upper bound that assumes full cooperation between the relay and the destination, and a constrained eavesdropper. The eavesdropper is constrained in the sense that it has to treat the relay's transmission as unknown noise for all subchannels, an idea used previously in the context of a class of classic relay-eavesdropper channel with orthogonal components in \cite{16}. These assumptions turn the parallel Gaussian relay-eavesdropper channel into a parallel Gaussian wiretap channel, the secrecy capacity of which is established in \cite{15}.

Furthermore, we apply the results developed for the parallel Gaussian relay-eavesdropper channel to the fading relay-eavesdropper channel, which is a special case of parallel Gaussian relay-eavesdropper channel, with each fading state corresponding to one subchannel. We illustrate our results through some numerical examples.

In this paper, the notation $X_{[1,L]}$ is used as a shorthand for $(X_1,X_2,\hdots,X_L)$, the notation $X_{[1,L]}^n$ is used as a shorthand for  $(X_1^n,X_2^n,\cdots,X_L^n)$ where for $l=1,\hdots,L$,  $X_l^n:=(X_{l1},X_{l2},\cdots,X_{ln})$ and the notation $\mc X_{1[1,L]}$ is used as a shorthand for $\mc X_{11}\times \mc X_{12}\hdots\times \mc X_{1L}$. We define the function $\mc{C}(x)$ as $\frac{1}{2}\log_2(1+x)$. Throughout the paper the logarithm function is taken to the base 2.
\section {The parallel Relay-Eavesdropper Channel}
A parallel relay-eavesdropper channel is a four terminal network consisting of  $\mc X_{1[1,L]}, \mc X_{2[1,L]}$ as finite input alphabets and $\mc Y_{[1,L]}, \mc Y_{1[1,L]}, \mc Y_{2[1,L]}$ as finite output alphabets. The transition probability distribution is given by
\begin{eqnarray}
\label{eq1}
\prod_{l=1}^L p(y_l,y_{1l},y_{2l} \mid x_{1l},x_{2l})
\end{eqnarray}
where $x_{1l} \in \mc X_{1l}, x_{2l} \in \mc X_{2l}, y_{1l} \in \mc Y_{1l}, y_{l} \in \mc Y_{l}$ and $y_{2l} \in \mc Y_{2l}$,  for $l= 1,\cdots,L$. For the subchannel $l$, $X_{1l}$ and $X_{2l}$ are the inputs from the source and relay; and $Y_{1l}, Y_{2l}, Y_l$ are the outputs at the relay, eavesdropper and  destination, respectively.
The source sends a message ${W} \in \mathcal{W} = \{1,\cdots,2^{n{{R}}} \}$ using a ($2^{nR},n$) code consisting of: 1) a stochastic encoder at the source that maps $W \rightarrow {X}_{1[1,L]}^n$, 2) a relay encoder that maps $f_i(Y_{1[1,L]}^{i-1})\rightarrow X_{2[1,L],i}$ for $1 \le i \le n$, and 3) a decoding function $g(Y_{[1,L]}^n) \rightarrow W$. The average error probability of a ($2^{nR},n$) code is defined as
\begin{eqnarray}
P_{e}^n = \frac{1}{2^{n{R}}} \sum_{W \in \mathcal{W}}p\{ g(Y_{[1,L]}^n) \ne W | W  \}.
\end{eqnarray}
The eavesdropper listens to what the source and relay transmit for free, due to the wireless nature of the medium. It then tries to guess the information being transmitted. Denoting  $Y_{2[1,L]}^n$ the output at the eavesdropper, the equivocation rate per channel use is defined as $R_e = H(W|Y_{2[1,L]}^n)/n$. Perfect secrecy for the channel is obtained when the eavesdropper gets no information about $W$ from $ Y_{2[1,L]}^n$. That is, the equivocation rate is equal to the unconditional source entropy. A rate equivocation pair (${{R},{R}_{e}}$) is achievable, if for any $\epsilon > 0$ there exists a sequence of codes ($2^{n{R}}$, $n$) such that for any $n \ge n(\epsilon)$
\begin{eqnarray}
\frac{H(W)}{n} &\ge& {R} - \epsilon,\notag \\
\frac{H(W| Y_{2[1,L]}^n)}{n} &\ge& {R_{e}} - \epsilon , \notag \\
P_{e}^n &\le& \epsilon.
\end{eqnarray}


\section{Discrete memoryless channel}
In this section, we establish outer and inner bounds on the rate-equivocation region of a parallel relay-eavesdropper channel.
\subsection{Outer bound}
\begin{theorem}
\label{Up}
For the parallel relay-eavesdropper channel with $L$ subchannels, and for any achievable rate-equivocation pair $(R,R_e)$, there exists a set of random variables ${U_l}\rightarrow({V_{1l},V_{2l}})\rightarrow({X_{1l},X_{2l}})\rightarrow({Y_l,Y_{1l},Y_{2l}})$, $l=1,\hdots,L$, such that ($R,R_e$) satisfies
\setlength{\arraycolsep}{0.2em}
\begin{eqnarray}
\label{Upeq}
{R}   &\le& \min \bigg \{ \sum_{l=1}^L {I({V_{1l}V_{2l}};Y_l)},\sum_{l=1}^{L} {I(V_{1l};Y_lY_{1l} \mid V_{2l})} \bigg\} \notag \\
{R}_e &\le& {R} \notag \\
{R}_e &\le& \min \bigg\{\sum_{l=1}^L I(V_{1l}V_{2l};Y_l\mid U_l)-{I(V_{1l}V_{2l};Y_{2l}\mid U_l)}, \notag \\&&\sum_{l=1}^L {I(V_{1l};Y_lY_{1l} \mid V_{2l}U_l)}-{I(V_{1l}V_{2l};Y_{2l}\mid U_l)}\bigg\}.
 \end{eqnarray}
 \setlength{\arraycolsep}{5pt}
\end{theorem}
\vspace{0.5em}
\begin{IEEEproof}
 The proof of Theorem 1 is given in Appendix \ref{app1}.
\end{IEEEproof}

\subsection{Achievable rate-equivocation region}
\begin{theorem}\label{low}
For the parallel relay-eavesdropper channel with $L$ subchannels, the rate pairs in the closure of the convex hull of all ($R, R_{e}$) satisfying
\setlength{\arraycolsep}{0.2em}
\begin{eqnarray}
\label{innerd}
R &\le&  \min \bigg \{\sum_{l\in \mc{A}}I(V_{1l}V_{2l};Y_l|U_l),\sum_{l\in \mc{A}} I(V_{1l};Y_{1l} | V_{2l}U_l)\bigg\} \notag \\
&&+\sum_{l\in \mc{A}^c} I(V_{1l};Y_l|V_{2l}) \notag \\
R_e &\le& R \notag \\
R_e &\le&  \min \bigg \{\sum_{l\in \mc{A}}  I(V_{1l}V_{2l};Y_l| U_l) -I(V_{1l}V_{2l};Y_{2l}| U_l),\notag \\ && \sum_{l\in \mc{A}} I(V_{1l};Y_{1l} | V_{2l}U_l)-I(V_{1l}V_{2l};Y_{2l}| U_l) \bigg\} \notag \\
&&+\sum_{l\in \mc{A}^c} I(V_{1l};Y_l | V_{2l})+ \min \bigg\{\sum_{l\in \mc{A}^c}I(V_{2l};Y_l),\notag\\&&\sum_{l\in \mc{A}^c} I(V_{2l};Y_{2l} | V_{1l})\bigg\}-\min \bigg \{\sum_{l\in \mc{A}^c}I(V_{2l};Y_l),\notag\\&& \sum_{l\in \mc{A}^c}I(V_{2l};Y_{2l})\bigg\}-\sum_{l\in \mc{A}^c} I(V_{1l};Y_{2l}| V_{2l}),
\end{eqnarray}
\end{theorem}
\setlength{\arraycolsep}{5pt}
\noindent for some distribution $p(u_l,v_{1l},v_{2l},x_{1l},x_{2l},y_{l},y_{1l},y_{2l})=p(u_l)p(v_{1l},v_{2l}|u_l)p(x_{1l},x_{2l}| v_{1l},v_{2l})p(y_{l},y_{1l},y_{2l}|x_{1l},x_{2l})$ for $l\in {\mc{A}}$ and $p(v_{1l},v_{2l},x_{1l},x_{2l},y_{l},y_{1l},y_{2l}) = p(v_{1l})p(v_{2l})p(x_{1l}|v_{1l})p(x_{2l}|v_{2l}) p(y_{l},y_{1l},y_{2l}|x_{1l},x_{2l})$ for $l \in {\mc{A}^c}$, are achievable.

\vspace{0.5em}
In the statement of Theorem 2, sets $\mc{A}$ and $\mc{A}^c$ represent the subchannels for which relay operates in DF and NF mode, respectively.
The region in Theorem \ref{low} is obtained through a coding scheme which combines appropriately DF and NF schemes. The rates for the DF scheme can be obtained readily by setting $U := U_{[1,|\mc {A}|]},V_{1}:= V_{1[1,|\mc {A}|]}, V_2 := V_{2[1,|\mc {A}|]},Y := Y_{[1,|\mc {A}|]}$, $Y_1 := Y_{1[1,|\mc {A}|]}$ and $Y_2 := Y_{2[1,|\mc {A}|]}$, for $l \in \mc{A}$ in \cite[Theorem 2]{10}. Similarly the rates for NF scheme can be readily obtained by setting $V_{1}:= V_{1[1,|\mc{A}^c|]}, V_2 := V_{2[1,|\mc{A}^c|]}, Y := Y_{[1,|\mc{A}^c|]}$, $Y_1 := Y_{1[1,|\mc{A}^c|]}$ and $Y_2 := Y_{2[1,|\mc{A}^c|]}$, for $l \in \mc{A}^c$ in \cite[Theorem 3]{10}.\\
\vspace{0.5em}
\begin{remark}
For a parallel relay-eavesdropper channel in which all subchannels are degraded{\footnote{In parallel relay-eavesdropper channel if all the subchannels  are degraded, the entire relay-eavesdropper channel may not necessarily be degraded.}}, i.e.,
\begin{multline}
 p(y_l,y_{1l},y_{2l} \mid x_{1l},x_{2l}) \\=  p(y_{1l}\mid x_{1l},x_{2l})p(y_{l}\mid y_{1l},x_{2l})p(y_{2l}\mid y_{l}, y_{1l}, x_{1l},x_{2l}),\notag
\end{multline}
for $l=1,\hdots,L$, the perfect secrecy capacity is given by
\setlength{\arraycolsep}{0.2em}
\begin{align}
\label{pd}
C_s =& \max \min \bigg\{\sum_{l=1}^L [I(V_{1l}V_{2l};Y_l\mid U_l)-{I(V_{1l}V_{2l};Y_{2l}\mid U_l)}]^+, \notag \\&\sum_{l=1}^L [{I(V_{1l};Y_{1l} \mid V_{2l}U_l)}-{I(V_{1l}V_{2l};Y_{2l}\mid U_l)}]^+\bigg\}
 \end{align}
where maximum is over ${U_l}\rightarrow({V_{1l},V_{2l}})\rightarrow({X_{1l},X_{2l}})\rightarrow({Y_l,Y_{1l},Y_{2l}})$, for $l=1,\hdots,L$.
 
\vspace{0.5em}
\begin{IEEEproof}
 The achievability follows from Theorem 2 by setting $\mc A^c=\varnothing$. The converse follows along the lines of Theorem 1 and is omitted for brevity.
\end{IEEEproof}
\end{remark}

\begin{figure*}[b]
 \hrulefill
\small{
\setcounter{equation}{9}
\begin{eqnarray}
\label{glow}
R_{e}^{\textrm{low}} =
{\max_{
\substack{
\sum_{l=1}^L P_{1l} \le P_1, \sum_{l=1}^L P_{2l} \le P_2, \\
0 \le\alpha_{l}\le 1,\hspace{.2em} \text{for $l=1,\hdots,|\mc{A}|$}}
}}
 \min \bigg \{ \sum_{l\in \mc{A}} \mathcal{C}\Bigg(
\frac{P_{1l}+\rho_{1l} P_{2l}+2\sqrt{\bar{\alpha}_l\rho_{1l}{P_{1l}P_{2l}}}}{{\sigma}_{l}^2}\Bigg)-\mathcal{C}\Bigg(
\frac{{P_{1l}+ \rho_{2l}P_{2l}}+2\sqrt{\bar{\alpha}_l\rho_{2l}{P_{1l}P_{2l}}}}{{\sigma}_{2l}^2}\Bigg),\notag\\ \sum_{l\in \mc{A}}\mathcal{C} \Bigg( \frac{\alpha_l{P_{1l}}}{\sigma_{1l}^2} \Bigg)-\mathcal{C}\Bigg(
\frac{{P_{1l}+ \rho_{2l}P_{2l}}+2\sqrt{\bar{\alpha}_l\rho_{2l}{P_{1l}P_{2l}}}}{{\sigma}_{2l}^2}\Bigg) \bigg \}
+\sum_{l \in \mc{A}^c} \mathcal{C} \Bigg( \frac{{P_{1l}}}{{\sigma}_{l}^2}\Bigg)+ \min \bigg \{ \sum_{l \in \mc{A}^c}\mathcal{C}\Bigg(\frac{\rho_{1l}P_{2l}}{P_{1l}+{\sigma}_{l}^2}\Bigg), \sum_{l \in \mc{A}^c}\mathcal{C}\Bigg(\frac{\rho_{2l}P_{2l}}{{\sigma}_{2l}^2}\Bigg) \bigg\}
\notag\\- \min \bigg \{\sum_{l \in \mc{A}^c} \mathcal{C}\Bigg(\frac{\rho_{1l}P_{2l}}{P_{1l}
+{\sigma}_{l}^2}\Bigg),\sum_{l \in \mc{A}^c} \mathcal{C}\Bigg(\frac{\rho_{2l}P_{2l}}{P_{1l}+{\sigma}_{2l}^2}\Bigg) \bigg\}- \sum_{l \in \mc{A}^c}\mathcal{C}\Bigg(\frac{P_{1l}}{{\sigma}_{2l}^2}\Bigg).
\end{eqnarray}}
\setlength{\arraycolsep}{0.001em}
 \vspace{.5mm}
\hrulefill
\small{
\begin{eqnarray}
\label{glowm}
{R}_{e}^{\textrm{low}} =
\max_{
\substack{
\sum_{l=1}^L P_{1l} \le P_1, \sum_{l=1}^L P_{2l} \le P_2, \\
0 \le\alpha_{l}\le 1,\hspace{.2em} \text{for $l=1,\hdots,|\mc{A}|$}}
}
 \min \bigg \{ \sum_{l\in \mc{A}} \bigg[\mathcal{C}\Bigg(
\frac{P_{1l}+\rho_{1l} P_{2l}+2\sqrt{\bar{\alpha}_l\rho_{1l}{P_{1l}P_{2l}}}}{{\sigma}_{l}^2}\Bigg) -\mathcal{C}\Bigg(
\frac{{P_{1l}+ \rho_{2l}P_{2l}}+2\sqrt{\bar{\alpha}_l\rho_{2l}{P_{1l}P_{2l}}}}{{\sigma}_{2l}^2}\Bigg)\bigg]^+ ,\notag\\\sum_{l\in \mc{A}}\bigg[\mathcal{C} \Bigg( \frac{\alpha_l{P_{1l}}}{\sigma_{1l}^2} \Bigg)-\mathcal{C}\Bigg(
\frac{{P_{1l}+ \rho_{2l}P_{2l}}+2\sqrt{\bar{\alpha}_l\rho_{2l}{P_{1l}P_{2l}}}}{{\sigma}_{2l}^2}\Bigg)\bigg]^+ \bigg \}
 \notag\\+ \min\Bigg \{\sum_{l \in \mc{A}^c}\bigg[\mathcal{C} \Bigg( \frac{P_{1l}+\rho_{1l}P_{2l}}{{\sigma}_{l}^2}\Bigg)-\mathcal{C}\Bigg(\frac{P_{1l}+\rho_{2l}P_{2l}}{{\sigma}_{2l}^2}\Bigg)\bigg]^+, \sum_{l \in \mc{A}^c}\bigg[\mathcal{C}\Bigg(\frac{P_{1l}}{{\sigma}_{l}^2}\Bigg)+\mathcal{C}\Bigg(\frac{\rho_{2l}P_{2l}}{{\sigma}_{2l}^2}\Bigg)-\mathcal{C}\Bigg(\frac{P_{1l}+\rho_{2l}P_{2l}}{{\sigma}_{2l}^2}\Bigg)\bigg]^+\Bigg\}.
\end{eqnarray}}
 \vspace{-5mm}
 \end{figure*}
\setlength{\arraycolsep}{5pt}
\normalsize
\section{Gaussian channel}
In this section we study a  parallel Gaussian relay-eavesdropper channel. We focus on perfectly secure achievable rates, i.e., $(R,R_e)$= $(R,R)$. The received signals at the relay, destination and eavesdropper are given by%
\setcounter{equation}{6}
\begin{align}
\label{gchan}
{Y_{1l,i}}&= {X_{1l,i}}+{Z_{1l,i}}\notag\\
{Y_{l,i}} &= {X_{1l,i}}+\sqrt{\rho_{1l}}{X_{2l,i}}+{Z_{l,i}}\notag\\
{Y_{2l,i}}&= {X_{1l,i}}+\sqrt{\rho_{2l}}{X_{2l,i}}+{Z_{2l,i}}
\end{align}
where $i$ is the time index, $\{Z_{1l,i}\},\{Z_{l,i}\}$ and $\{Z_{2l,i}\}$ are noise processes, independent and identically distributed (i.i.d) with the components being zero mean Gaussian random variables with variances $\sigma_{1l}^2$, $\sigma_{l}^2$ and $\sigma_{2l}^2$; $X_{1l,i}$ and $X_{2l,i}$ are the inputs from the source and relay nodes respectively. The parameter $\rho_{1l}$ indicates the ratio of the R-D link signal-to-noise (SNR) to the S-D link SNR and $\rho_{2l}$ indicates the ratio of the R-E link SNR to the S-E link SNR for the $l^{th}$ subchannel. The source and relay input sequences are subject to the following average transmit power constraints
\begin{eqnarray}
\label{p_con1}
\frac{1}{n} \sum_{l=1}^L\sum_{i=1}^n \mathbb{E}[X_{1l,i}^2]\le  P_1, \\
 \label{p_con2}
 \frac{1}{n}  \sum_{l=1}^L \sum_{i=1}^n \mathbb{E}[X_{2l,i}^2]\le P_2.
\end{eqnarray}

\subsection{Lower bound on the perfect secrecy rate}
\label{section}
For the  parallel Gaussian relay-eavesdropper channel, defined by \eqref{gchan}, we apply Theorem \ref{low} to obtain a lower bound on the perfect secrecy rate.
\vspace{0.5em}
\begin{corollary}
For the  parallel Gaussian relay-eavesdropper channel (\ref{gchan}), a lower bound on the perfect secrecy rate is given by \eqref{glow}.
\end{corollary}
\vspace{0.5em}
\begin{IEEEproof}
 The achievability follows by applying Theorem \ref{low} with the choice $U_l$ = constant, $V_{1l}$=$X_{1l}$, $V_{2l}$=$X_{2l}$, ${X_{1l}} = {\tilde{X}_{1l}} + \sqrt{\frac{\bar{\alpha}_lP_{1l}}{P_{2l}}} {X_{2l}}$, $\bar{\alpha}_l=1-\alpha_l$, $\tilde{X}_{1l}\sim\mc N(0,\alpha_lP_{1l})$ independent of $ X_{2l}\sim\mc N(0,P_{2l})$ for $l \in {\mc{A}}$; and $X_{1l}\sim\mc N(0,P_{1l})$ independent of $X_{2l}\sim\mc N(0,P_{2l})$ for $l\in\mc{A}^c$. Straightforward algebra which is omitted for brevity gives \eqref{glow}.
\end{IEEEproof}
\vspace{0.5em}
The parameters $P_{1l}$ and $P_{2l}$ indicate the source and relay power allocated for transmission over the subchannel $l$. In \eqref{glow}, after some straightforward algebra, the contribution to the equivocation of information sent through NF (set $\mc{A}^c$) can be condensed by observing that we only need to consider  $\min\{\sum_{l \in \mc{A}^c}I(X_{2l},Y_{2l}),\sum_{l \in \mc{A}^c}I(X_{2l},Y_{l})\}=\sum_{l \in \mc{A}^c}I(X_{2l},Y_{2l})$ in set $\mc{A}^c$, to get a higher secrecy rate. A simplified expression for $R_e^{\textrm{low}}$ is given by \eqref{glowm}.
\subsection{Upper bound on the perfect secrecy rate}
The following theorem provides an upper bound on the secrecy rate of the parallel Gaussian relay-eavesdropper channel.
\vspace{0.1em}
\begin{theorem}
For the parallel Gaussian relay-eavesdropper channel \eqref{gchan}, an upper bound on the secrecy rate is given by
\setcounter{equation}{11}
\begin{eqnarray}
\label{guu}
{R}_e^{\text{up}} \le \max_{\{{\bf{K}_{P_\textit{l}}} \in {{\mc{K}_{P_\textit{l}}}}\}_{\textit{l}=1\hdots L}}  \sum_{l=1}^L I(X_{1l}X_{2l};Y_l)-I(X_{1l}X_{2l};Y_{2l})
\end{eqnarray}
where the maximization is over $[X_{1l},X_{2l}] \sim \mc{N}(\bf{0},\bf{K}_{P_\textit{l}})$ with ${{\mc{K}}_{P_\textit{l}}}  = \bigg \{ {\bf{K}_{P_\textit{l}}} : {\bf{K}_{P_\textit{l}}}=  \left [
 \begin{smallmatrix}
  P_{1l} & \psi_\textit{l }\sqrt{P_{1l} P_{2l}}\\
\psi_\textit{l }\sqrt{P_{1l} P_{2l}} & P_{2l}
 \end{smallmatrix} \right ], -1\le \psi_\textit{l} \le {1} \bigg\}$, for $l=1,\hdots,L$, with the covariance matrices $\mathbb{E}[X_{1[1,L]}X_{1[1,L]}^T]$,  $\mathbb{E}[X_{2[1,L]}X_{2[1,L]}^T]$ satisfying \eqref{p_con1} and \eqref{p_con2} respectively.
\setlength{\arraycolsep}{0.2em}
 \setlength{\arraycolsep}{5pt}
\end{theorem}
\vspace{0.5em}

\begin{IEEEproof}
The result in Theorem 1 established for the DM case can be extended to memoryless channels with discrete time and continuous alphabets using standard techniques \cite[Chapter 7]{22}. Taking the first term of the minimization in the bound on the equivocation rate, we get
\begin{eqnarray}
\label{nupper}
R_{e}\le  \max \sum_{l=1}^L  I(V_{1l}V_{2l} ;Y_{l}\mid U_{l})-I(V_{1l}V_{2l} ;Y_{2l}\mid U_{l})
\end{eqnarray}
where ${U_l}\rightarrow({V_{1l},V_{2l}})\rightarrow({X_{1l},X_{2l}})\rightarrow({Y_l,Y_{1l},Y_{2l}})$, for $l=1,\hdots,L$. The rest of the proof uses elements from related works in \cite{15} and \cite{19}. Continuing from \eqref{nupper}, we obtain
\setlength{\arraycolsep}{.2pt}
\begin{eqnarray}
R_e &\leq& \sum_{l=1}^L  I(V_{1l}V_{2l} ;Y_{l}\mid U_{l})-I(V_{1l}V_{2l} ;Y_{2l}\mid U_{l})\notag\\
  &\overset{(a)}{\le}& \sum_{l=1}^L I(V_{1l} V_{2l};Y_l )-I(V_{1l}V_{2l} ;Y_{2l})\notag \\
  &{\le}& \sum_{l=1}^L I(V_{1l} V_{2l};Y_l Y_{2l} )-I(V_{1l}V_{2l} ;Y_{2l})\notag \\
   &\overset{(b)}{=}& \sum_{l=1}^L [I( X_{1l} X_{2l};Y_l Y_{2l} )-I( X_{1l} X_{2l};Y_l Y_{2l}\mid V_{1l} V_{2l} )] \notag\\
   &&- [I(X_{1l} X_{2l} ;Y_{2l})-I(X_{1l} X_{2l} ;Y_{2l}\mid V_{1l}V_{2l} )]\notag \\
   &{=}& \sum_{l=1}^L [I( X_{1l} X_{2l};Y_l Y_{2l} )-I(X_{1l} X_{2l} ;Y_{2l})] \notag\\&
     &- [I( X_{1l} X_{2l};Y_l Y_{2l}\mid V_{1l} V_{2l} )-I(X_{1l} X_{2l} ;Y_{2l}\mid V_{1l}V_{2l} )]\notag \\
     &{\le}& \sum_{l=1}^L [I( X_{1l} X_{2l};Y_l Y_{2l})-I(X_{1l} X_{2l} ;Y_{2l})] \notag\\
     &=&  \sum_{l=1}^L I( X_{1l} X_{2l};Y_l \mid Y_{2l}),
     \label{ubound}
     \end{eqnarray}   
\noindent where $(a)$ follows  by noticing that $I(V_{1l}V_{2l} ;Y_{l}\mid U_{l})-I(V_{1l}V_{2l} ;Y_{2l}\mid U_{l})$ is maximized by setting $U_l$=constant and $(b)$ follows from the Markov chain condition $({V_{1l},V_{2l}})\rightarrow({X_{1l},X_{2l}})\rightarrow({Y_l,Y_{1l},Y_{2l}})$, for $l =1,\hdots,L$.

We now tighten the upper bound \eqref{ubound} by using an argument previously used in \cite{19}, \cite{21} in the context of multi-antennas wiretap channels. More specifically, observing that, the original bound \eqref{nupper} depends on $p(y_{l},y_{2l}|x_{1l},x_{2l})$ only through its marginals $p(y_{l}|x_{1l},x_{2l})$ and $p(y_{2l}|x_{1l},x_{2l})$, the upper bound \eqref{ubound} can be further tightened as
  \setlength{\arraycolsep}{5pt}
\begin{eqnarray}
\label{nuppern}
R_{e}\le   \max_{p(x_{1l},x_{2l})} \sum_{l=1}^L \min_{p(y'_{l},y'_{2l}|x_{1l},x_{2l})} I( X_{1l} X_{2l};Y'_l \mid Y'_{2l})
\end{eqnarray}
where the joint conditional $p(y'_{l},y'_{2l}|x_{1l},x_{2l})$ has the same marginals  as $p(y_{l},y_{2l}|x_{1l},x_{2l})$, i.e.,  $p(y'_{l}|x_{1l},x_{2l})=p(y_{l}|x_{1l},x_{2l})$ and $p(y'_{2l}|x_{1l},x_{2l})=p(y_{2l}|x_{1l},x_{2l})$.

It can be easily shown that the bound in \eqref{nuppern} is maximized when the inputs are jointly Gaussian, i.e.,  $[X_{1l},X_{2l}] \sim \mc{N}(\bf{0},\bf{K}_{P_\textit{l}})$, ${\bf{K}_{P_\textit{l}} \in \mc{K}_{P_\textit{l}}}$ with $\mc{K}_{P_\textit{l}} = \bigg \{ {\bf{K}_{P_\textit{l}}} : {\bf{K}_{P_\textit{l}}}=  \left [
 \begin{smallmatrix}
  P_{1l} & \psi_\textit{l }\sqrt{P_{1l} P_{2l}}\\
\psi_\textit{l }\sqrt{P_{1l} P_{2l}} & P_{2l}
 \end{smallmatrix} \right ], -1\le \psi_\textit{l} \le {1} \bigg\}$, for $l=1,\hdots,L$ with the covariance matrices $\mathbb{E}[X_{1[1,L]}X_{1[1,L]}^T]$ and $\mathbb{E}[X_{2[1,L]}X_{2[1,L]}^T]$ satisfying \eqref{p_con1} and \eqref{p_con2} respectively \cite{19},\cite{21}.
 
Next, using the specified Gaussian inputs, it can be shown that the evaluation of the upper bound \eqref{nuppern} minimized over all possible correlations between $Y'_{l}, Y'_{2l}$, for $l=1,\hdots,L$ yields
 \begin{eqnarray}
 \label{lab1}
 R_e \le  \max_{\{{\bf{K}_{P_\textit{l}}} \in {{\mc{K}_{P_\textit{l}}}}\}_{\textit{l}=1\hdots L}}\sum_{l=1}^L  I( X_{1l} X_{2l};Y_l)-  I( X_{1l} X_{2l}; Y_{2l}).
  \label{ubm}
     \end{eqnarray}
 This concludes the proof.
 \end{IEEEproof}
\vspace{0.5em}
\noindent We now study the case in which the links S-R are very noisy, i.e., the relay does not hear the source.
\vspace{0.5em}
\begin{theorem}
For the model \eqref{gchan} in which the relay does not hear the source, the secrecy capacity is given by
\setcounter{equation}{16}
\begin{multline}
\label{eqbound}
C_s =  \min \bigg\{
\max \sum_{l=1}^L \mc{C} \bigg(\frac{P_{1l}+\rho_{1l}P_{2l}}{\sigma_{l}^2}\bigg)- \mc{C} \bigg(\frac{P_{1l}+\rho_{2l}P_{2l}}{\sigma_{2l}^2}\bigg),\\
\max \sum_{l=1}^L \mc{C} \bigg(\frac{P_{1l}}{\sigma_{l}^2}\bigg)- \mc{C} \bigg(\frac{P_{1l}}{\sigma_{2l}^2+\rho_{2l}P_{2l}}\bigg) \bigg \}
\end{multline}
where the maximization is over $\{P_{1l},P_{2l}\}$, for $l=1\hdots L$, such that $\sum_{l=1}^L P_{1l} \leq P_1$ and $\sum_{l=1}^L P_{2l} \leq P_2$.
\setlength{\arraycolsep}{0.2em}
 \setlength{\arraycolsep}{5pt}
\end{theorem}
\vspace{0.5em}
\begin{IEEEproof}\\
\textbf{Upper Bound}:
The bound given by the first term of the minimization in \eqref{eqbound} follows from a straightforward application of the result in Theorem 3 --- taking independent source and relay inputs since the relay does not hear the source transmission in this case.

The bound given by the second term of the minimization in \eqref{eqbound} can be established as follows. Our approach borrows elements from an upper bounding technique that is used in \cite{16}, and can be seen as an extension of it to the case of parallel relay-eavesdropper channels. Assume that all the links between the relay and the destination are noiseless, and the eavesdropper is constrained to treat the relay's signal as unknown noise. As mentioned in \cite{16}, any upper bound for this model with full relay-destination cooperation and constrained eavesdropper, also applies for the general model.

Now, for the model with full relay-destination cooperation and constrained eavesdropper, we develop an upper bound on the secrecy capacity as follows. In this case, the destination can remove the effect of the relay transmission (which is independent from the source transmission as the relay does not hear the source), and the equivalent channel to the destination can be written as
\begin{align}
{Y'_{l,i}} &= {X_{1l,i}}+{Z_{l,i}}.
\end{align}
For the constrained eavesdropper the relay's transmission acts as an interference, with the worst case obtained with Gaussian $X_{2[1,L]}$ \cite{16}. The equivalent output at the eavesdropper in this case is given by
\begin{align}
{Y'_{2l,i}}&= {X_{1l,i}}+\sqrt{\rho_{2l}\mathbb{E}[{X^2_{2l,i}}]}+{Z_{2l,i}}.
\end{align}
The rest of the proof follows by simply observing that the resulting model (with the worst case relay transmission to the eavesdropper and full relay-destination cooperation) is, in fact a parallel Gaussian wiretap channel, the secrecy capacity of which is established in \cite{15},
\begin{eqnarray}
\label{newup2}
C_s \le  \max \sum_{l=1}^L  I(X_{1l};Y'_{l})-I(X_{1l};Y'_{2l})
\end{eqnarray}
where the maximization is over $X_{1l}\sim \mc{N}(0,P_{1l}),X_{2l} \sim \mc{N}(0,P_{2l})$ for $l=1\hdots L$, with $\sum_{l=1}^{L}P_{1l} \leq P_1$ and $\sum_{l=1}^{L}P_{2l} \leq P_2$.

Finally straightforwad algebra which is omitted for brevity shows that the computation of \eqref{newup2} gives the second term of the minimization in \eqref{eqbound}.
\vspace{0.5em}
\\\textbf{Lower Bound}:
The achievability follows by computing the lower bound in Theorem 2 with the choices $|\mc{A}^c|$:=$L$, $V_{1l}$:=$X_{1l}$, $V_{2l}$:=$X_{2l}$, and  $X_{1l}\sim\mc N(0,P_{1l})$ independent of $X_{2l}\sim\mc N(0,P_{2l})$.
\end{IEEEproof}

\section{Fading relay-eavesdropper channel}
We consider a fading relay-eavesdropper channel that is corrupted by multiplicative fading gain processes in addition to additive white Gaussian noise (AWGN) processes. The received signals are given by
  \begin{align}
\label{raych}
{Y_{1,i}}&= {h_{sr,i}{X_{1,i}}}+{Z_{1,i}}\notag\\
{Y_{i}}  &= {h_{sd,i}X_{1,i}}+{h_{rd,i}X_{2,i}}+{Z_{i}}\notag\\
{Y_{2,i}}&= {h_{se,i}X_{1,i}}+{h_{re,i}X_{2,i}}+{Z_{2,i}}
\end{align}
where $i$ is the time index, $ h_{sd,i}$, $h_{rd,i}$, $h_{se,i}$, $h_{re,i}$ and  $h_{sr,i}$ are the fading gain coefficients associated with S-D, R-D, S-E, R-E and S-R links, given by complex Gaussian random variables with zero mean and unit variance respectively. The noise processes $\{Z_{1,i}\},\{Z_{i}\},\{Z_{2,i}\}$ are zero mean i.i.d complex Gaussian random variables with  variances $\sigma_{1}^2$, $\sigma^2$ and $\sigma_{2}^2$ respectively. The source and relay input sequences are subject to an average power constraint, i.e, $\sum_{i=1}^n \mathbb{E}[|X_{1,i}|^2] \le nP_1$, $\sum_{i=1}^n \mathbb{E}[|X_{2,i}|^2] \le nP_2$. Let $\bar{h}_i := [h_{sd,i} \hspace{.5em} h_{rd,i} \hspace{.5em}  h_{se,i} \hspace{.5em}   h_{re,i} \hspace{.5em}  h_{sr,i}]$ and we assume that  perfect channel state information (CSI) is available at all nodes, i.e, each node has access to the instantaneous CSI and its statistics. For a given fading state realization $\bar{h}_i$, the fading relay-eavesdropper channel is a Gaussian relay-eavesdropper channel. Therefore, for a given channel state with $L$ fading state realizations,  the fading relay-eavesdropper channel can be seen as a parallel Gaussian relay-eavesdropper channel with $L$ subchannels. The power allocation vectors at the source and relay are denoted by $P_1(\bar{h})$ and $P_2(\bar{h})$ respectively. The ergodic achievable secrecy rate of the fading relay-eavesdropper channel \eqref{raych}, which follows from \eqref{glowm} is given by \eqref{fcap}. The upper bound for the fading  relay-eavesdropper channel follows from \eqref{guu} and is given by \eqref{eqfg}. In the achievable region we proposed a coding scheme which is a combination of DF and NF scheme. A pertinent question is how to decide which scheme to use on each subchannel ? To accomplish this we define $\mc{A}$:= $\{\bar{h} : |{h}_{sd}|^2 < |{h}_{sr}|^2\}$  contains all the fading state realizations in $\{\bar{h}\}$ where the S-R link is better than S-D link. The complement of set $\mc{A}$ is  $\mc{A}${$^c$}:= $\{\bar{h} : |{h}_{sd}|^2\ge|{h}_{sr}|^2\}$. 
 
\section{Numerical Results}
We consider a fading relay-eavesdropper channel with \textit{L} fading states. It is assumed that perfect channel state information is available at all nodes. We can consider this channel as a Gaussian relay-eavesdropper channel with \textit{L} subchannels. We model channel gain between node $i \in \{s,r\}$ and $j \in \{r,d,e\}$ as distance dependent Rayleigh fading, that is, $h_{i,j} = h'_{i,j}{d_{i,j}^{-\gamma/2}}$,
where $\gamma$ is the path loss exponent and $h'_{i,j}$ is a complex Gaussian random variable with zero mean and variance one. Each subchannel is corrupted by AWGN with zero mean and variance one. The objective function for both lower and upper bounds  are optimized numerically using AMPL with a commercially available solver, for instance SNOPT. Furthermore, for each symbol transmission same subchannel is used on S-R and R-D links to make the optimization tractable. 
\begin{figure*}[t]
 \setcounter{equation}{21}
  \small
\begin{align}
\label{fcap}
{R}_{e}^{\textrm{low}} =&
\max_{
\substack{
\mathbb{E}[P_1(\bar{h})] \le P_1,\\ \mathbb{E}[P_2(\bar{h})] \le P_2,\\
0 \le\alpha(\bar{h})\le 1}} \min \bigg \{ \mathbb{E}_{\bar{h}\in \mc{A}} \bigg[2\mathcal{C}\Bigg(
\frac{|{h}_{sd}|^2P_1(\bar{h})+ |{h}_{rd}|^2P_2(\bar{h})+2\sqrt{\bar{\alpha}(\bar{h})|{h}_{sd}|^2P_1(\bar{h})|h_{rd}|^2P_2(\bar{h})}}{{\sigma}^2}\Bigg)\notag\\&-2\mathcal{C}\Bigg(\frac{|{h}_{se}|^2P_1(\bar{h})+|{h}_{re}|^2P_2(\bar{h})+2\sqrt{\bar{\alpha}(\bar{h})|{h}_{se}|^2P_1(\bar{h})|{h}_{re}|^2P_2(\bar{h})}}{{\sigma}_{2}^2}\Bigg)\bigg]^+, \mathbb{E}_{\bar{h} \in \mc{A}} \bigg[2\mathcal{C} \Bigg( \frac{\alpha(\bar{h})|{h}_{sr}|^2P_1(\bar{h})}{\sigma_{1}^2}\Bigg)\notag\\&-2\mathcal{C}\Bigg(\frac{|{h}_{se}|^2P_1(\bar{h})+|{h}_{re}|^2P_2(\bar{h})+2\sqrt{\bar{\alpha}(\bar{h})|{h}_{se}|^2P_1(\bar{h})|{h}_{re}|^2P_2(\bar{h})}}{{\sigma}_{2}^2}\Bigg)\bigg]^+\bigg\} + \min\bigg\{ \mathbb{E}_{\bar{h} \in \mc{A}^c} \bigg [2\mathcal{C} \Bigg( \frac{|{h}_{sd}|^2{P_1(\bar{h})}+|{h}_{rd}|^2{P_2(\bar{h})}}{\sigma^2}\Bigg)\notag\\&-2\mathcal{C} \Bigg( \frac{|{h}_{se}|^2{P_1(\bar{h})}+|{h}_{re}|^2{P_2(\bar{h})}}{\sigma_{2}^2}\Bigg)\bigg]^+,\mathbb{E}_{\bar{h} \in \mc{A}^c} \bigg[ 2\mathcal{C}\Bigg(\frac{|{h}_{sd}|^2P_1(\bar{h})}{\sigma^2}\Bigg) + 2\mathcal{C}\Bigg(\frac{|{h}_{re}|^2P_2(\bar{h})}{\sigma_{2}^2}\Bigg)-  2\mathcal{C} \Bigg( \frac{|{h}_{se}|^2{P_1(\bar{h})}+|{h}_{re}|^2{P_2(\bar{h})}}{{\sigma}_{2}^2}\Bigg)\bigg ]^+\bigg \}.
\end{align}
 \hrulefill
\vspace*{-5mm}
 \end{figure*}
\begin{figure*}[!t]
\setcounter{equation}{22}
\small {
\begin{align}
\label{eqfg}
R_e^{\text{up}} \le\max_{
\substack{
\mathbb{E}[P_1(\bar{h})] \le P_1,\\ \mathbb{E}[P_2(\bar{h})] \le P_2,\\
-1 \le \psi(\bar{h}) \le 1}} &\mathbb{E}_{\bar{h}} \bigg \{
2\mc{C}\bigg(\frac{|h_{sd}|^2 P_{1}(\bar{h})+|h_{rd}|^2P_{2}(\bar{h})+2\psi(\bar{h}) \sqrt{|h_{sd}|^2P_{1}(\bar{h})|h_{rd}|^2P_{2}(\bar{h})}}{\sigma^2}\bigg)\notag\\&- 2\mc{C}\bigg(\frac{|h_{se}|^2 P_{1}(\bar{h})+|h_{re}|^2P_{2}(\bar{h})+2\psi(\bar{h}) \sqrt{|h_{se}|^2P_{1}(\bar{h})|h_{re}|^2P_{2}(\bar{h})}}{\sigma_{2}^2}\bigg) \bigg\}.
\end{align}
 }
 \hrulefill
 \vspace{-2mm}
\end{figure*}

\begin{figure}[ht]
\centering
\includegraphics[width=9cm, height=6cm]{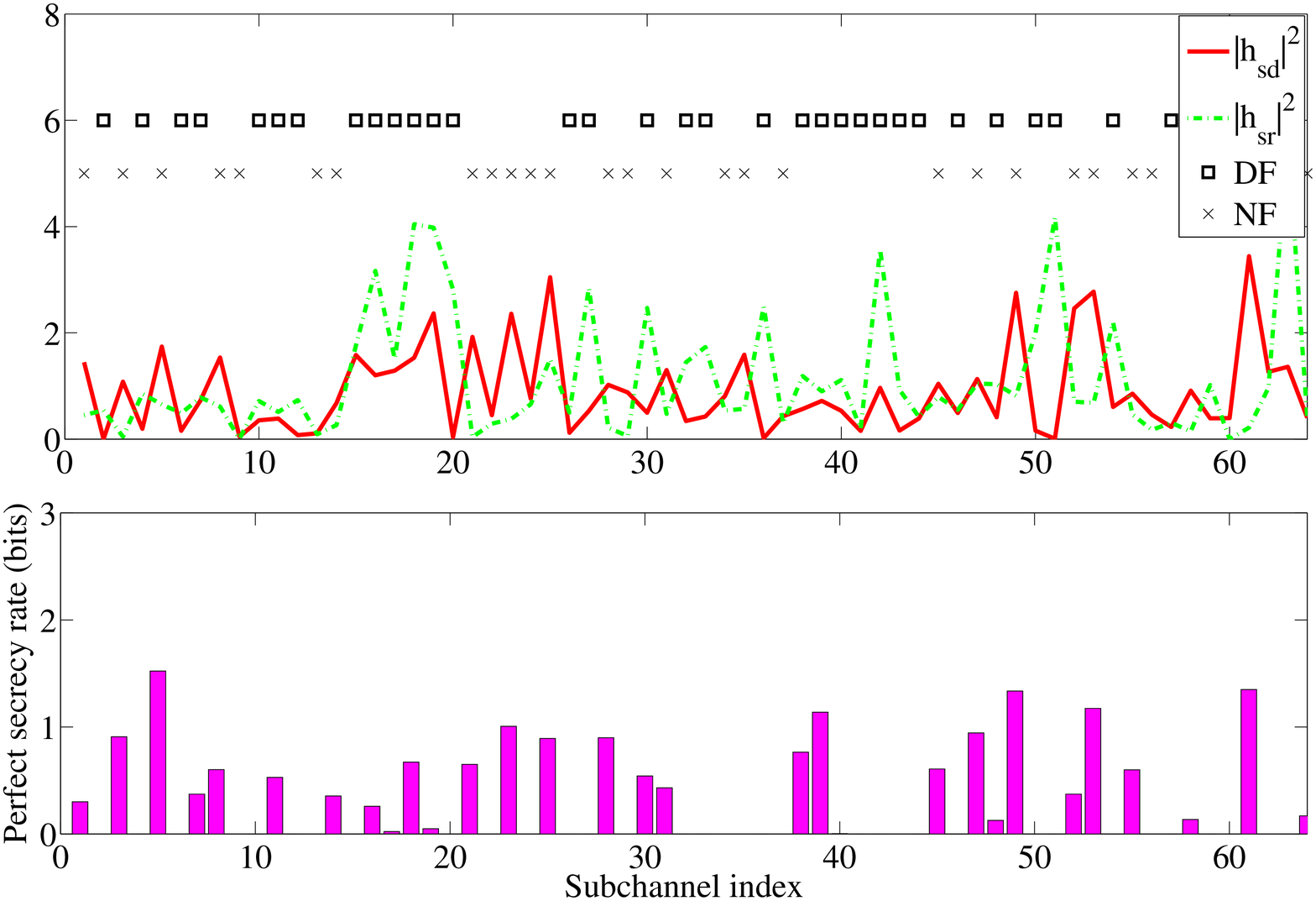}
 \caption{Achievable perfect secrecy rate of a fading parallel relay-eavesdropper channel.}\label{pcp}
\end{figure}

To illustrate the system performance, we set the source and relay power to 64 Watt each. We consider a network geometry in which the source is located at the point (0,0), the relay is located at the point ($d$,0), the destination is located at the point (1,0) and the eavesdropper is located at the point (0,1), where $d$ is the distance between the source and the relay. In all numerical results we set path loss exponent $\gamma$:=2.
 Fig. \ref{pcp} shows the power allocation for a fading channel with 64 subchannels  where the relay is located at (0.5,0), and  marker `$\times$' denotes NF on a particular subchannel while marker `$\square$' denotes DF on a particular subchannel. It can be seen from Fig. \ref{pcp} that, achievable perfect secrecy rate is zero for some subchannels. Roughly speaking, this happens when the condition $|h_{rd}|^2>|h_{re}|^2$  is violated.

 \begin{figure}[ht]
\centering
\includegraphics[width=\linewidth]{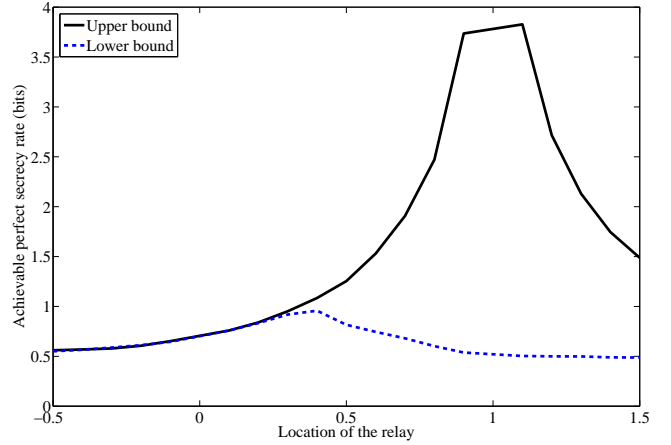}
\caption{Bounds on perfect secrecy rate.}
 \label{fig6}
 \end{figure}
 
Mode selection at the relay by only considering the relative strength of the S-D and the S-R link is suboptimal because the achievable secrecy rate \eqref{fcap} also depends on the other link gain. We now consider the case in which the relay independently selects the scheme which maximize the rate for each subchannel.  When the relay is close to the source, it uses DF scheme on all subchannels. Similarly when the relay is close to the destination, use of NF scheme on all subchannels offers better rate. The region when the relay is between $0.5 < d < 1.2$, it selects between DF and NF scheme based on link gains of S-D, S-R link as mentioned above. In Fig. \ref{fig6} we plot the optimized lower and upper bounds on the secrecy rate for fading relay-eavesdropper channel. It can be seen that when the relay is close to the source the lower and upper bound matches. This follows because of using DF scheme on all subchannels.


\appendices
\section{}
\label{app1}
The proof generalizes the results of Theorem 1 in \cite{10} and uses elements from a similar proof in the context of parallel BCC in \cite{15}.

\vspace{1em}
\textbf{a)} We first bound the equivocation rate as follows.
\setcounter{equation}{23}
\setlength{\arraycolsep}{0.2em}
\begin{eqnarray}
\label{upper}
nR_{e}&=& {H}(W \mid Y_{2[1,L]}^n)\notag \\
&=& H({W})- I(W;Y_{2[1,L]}^n)\notag\\
&=& I(W;Y_{[1,L]}^n)-I(W;Y_{2[1,L]}^n)+{H}(W \mid Y_{[1,L]}^n)\notag \\
&\overset{(a)}{\le}& I(W;Y_{[1,L]}^n)-I(W;Y_{2[1,L]}^n)+n\epsilon_{n} \notag\\
&=& \sum_{l=1}^L I(W;Y_l^n\mid Y_{[1,l-1]}^n)-I(W;Y_{2l}^n\mid Y_{2[l+1,L]}^n)+n\epsilon_{n} \notag\\
&{=}& \sum_{l=1}^L \sum_{i=1}^n I(W;Y_{li}\mid Y_l^{i-1}Y_{[1,l-1]}^n)\notag\\&&- I(W;Y_{2li}\mid Y_{2l[i+1]}^n Y_{2[l+1,L]}^n)+n\epsilon_{n}\notag\\
&{=}& \sum_{l=1}^L \sum_{i=1}^n I(W Y_{2l[i+1]}^n Y_{2[l+1,L]}^n;Y_{li}\mid Y_l^{i-1}Y_{[1,l-1]}^n)\notag\\&&- I(Y_{2l[i+1]}^n Y_{2[l+1,L]}^n;Y_{li}\mid W Y_l^{i-1}Y_{[1,l-1]}^n)\notag\\&&- I(W Y_l^{i-1}Y_{[1,l-1]}^n;Y_{2li}\mid Y_{2l[i+1]}^n Y_{2[l+1,L]}^n)\notag\\&&+ I(Y_l^{i-1}Y_{[1,l-1]}^n;Y_{2li}\mid W Y_{2l[i+1]}^n Y_{2[l+1,L]}^n)+n\epsilon_{n}\notag \\
&\overset{(b)}{\le}& \sum_{l=1}^L \sum_{i=1}^n I(W Y_{2l[i+1]}^n Y_{2[l+1,L]}^n;Y_{li}\mid Y_l^{i-1}Y_{[1,l-1]}^n)\notag\\&&- I(W Y_l^{i-1}Y_{[1,l-1]}^n;Y_{2li}\mid Y_{2l[i+1]}^n Y_{2[l+1,L]}^n)+n\epsilon_{n}\notag \\
&=& \sum_{l=1}^L \sum_{i=1}^n I(Y_{2l[i+1]}^n Y_{2[l+1,L]}^n;Y_{li}\mid Y_l^{i-1}Y_{[1,l-1]}^n) \notag\\&&+I(W ;Y_{li}\mid  Y_l^{i-1}Y_{[1,l-1]}^n Y_{2l[i+1]}^n Y_{2[l+1,L]}^n)\notag\\&&- I(Y_l^{i-1}Y_{[1,l-1]}^n;Y_{2li}\mid Y_{2l[i+1]}^n Y_{2[l+1,L]}^n)\notag\\&&-I(W;Y_{2li}\mid Y_l^{i-1}Y_{[1,l-1]}^n Y_{2l[i+1]}^n Y_{2[l+1,L]}^n)+n\epsilon_{n}\notag \\
&\overset{(c)}{=}& \sum_{l=1}^L \sum_{i=1}^n I(W ;Y_{li}\mid  Y_l^{i-1}Y_{[1,l-1]}^n Y_{2l[i+1]}^n Y_{2[l+1,L]}^n) \notag\\&&-I(W;Y_{2li}\mid Y_l^{i-1}Y_{[1,l-1]}^n Y_{2l[i+1]}^n Y_{2[l+1,L]}^n)+n\epsilon_{n}
\end{eqnarray}
where $\epsilon_n \rightarrow 0$ as $n \rightarrow \infty$;
$(a)$ follows from Fano's inequality; and
$(b)$ and $(c)$ follows from lemma 7 in \cite{9}.
 
We introduce a random variable $T$ uniformly distributed over $\{1,2,\cdots,n\}$ and set, $U_{li} =  Y_l^{i-1}Y_{[1,l-1]}^n Y_{2l[i+1]}^n Y_{2[l+1,L]}^n$, $V_{1li} =  W Y_{2l[i+1]}^n Y_{2[l+1,L]}^n$ and $V_{2li} =  Y_l^{i-1}Y_{[1,l-1]}^n$. We define  $U_{l} = (T,U_{li}), V_{1l} = (T,V_{1li}), V_{2l} = (T,V_{2li}), X_{1l}=X_{1T}, X_{2l}=X_{2T},Y_{l}=Y_{T}, Y_{1l}=Y_{1T}, Y_{2l}=Y_{2T}$, for $l=1,\cdots,L$.
Note that $(U_l,V_{1l},V_{2l},X_{1l},X_{2l},Y_{l},Y_{1l},Y_{2l})$ satisfies the following Markov Chain condition
\begin{equation}
{U_l}\rightarrow({V_{1l},V_{2l}})\rightarrow({X_{1l},X_{2l}})\rightarrow({Y_l,Y_{1l},Y_{2l}}),  \text{for $l =1,\cdots,L$}.\notag
\end{equation}
Thus, we have
\begin{eqnarray}
R_{e}&\le& \frac{1}{n}\sum_{l=1}^L \sum_{i=1}^n I(W ;Y_{li}\mid  Y_l^{i-1}Y_{[1,l-1]}^n Y_{2l[i+1]}^n Y_{2[l+1,L]}^n) \notag\\&&-I(W;Y_{2li}\mid Y_l^{i-1}Y_{[1,l-1]}^n Y_{2l[i+1]}^n Y_{2[l+1,L]}^n)+\epsilon_{n}\notag \\
&=& \frac{1}{n}\sum_{l=1}^L \sum_{i=1}^n I(W Y_l^{i-1}Y_{[1,l-1]}^n Y_{2l[i+1]}^n Y_{2[l+1,L]}^n ;Y_{li}\mid \notag\\&&  Y_l^{i-1}Y_{[1,l-1]}^n Y_{2l[i+1]}^n Y_{2[l+1,L]}^n) \notag \\&&-I(WY_l^{i-1}Y_{[1,l-1]}^n Y_{2l[i+1]}^n Y_{2[l+1,L]}^n;Y_{2li}\mid\notag \\&& Y_l^{i-1}Y_{[1,l-1]}^n Y_{2l[i+1]}^n Y_{2[l+1,L]}^n)+\epsilon_{n}\notag \\
\label{u1u}
&\overset{(d)}{=}&  \frac{1}{n} \sum_{l=1}^L \sum_{i=1}^n I(V_{1li}V_{2li} ;Y_{li}\mid U_{li})-I(V_{1li}V_{2li} ;Y_{2li}\mid U_{li})\notag\\&&+\epsilon_{n}\\
\label{u1}
&\overset{(e)}{\le}& \sum_{l=1}^L I(V_{1l}V_{2l} ;Y_{l}\mid U_{l})-I(V_{1l}V_{2l} ;Y_{2l}\mid U_{l})+\epsilon_{n}
\end{eqnarray}
where $(d)$ and $(e)$ follow by using the above definition.\\
We can also bound the equivocation rate as follows.
We continue from \eqref{upper} to get
\begin{eqnarray}
R_{e}&\le& \frac{1}{n}\sum_{l=1}^L \sum_{i=1}^n I(W ;Y_{li}\mid  Y_l^{i-1}Y_{[1,l-1]}^n Y_{2l[i+1]}^n Y_{2[l+1,L]}^n) \notag\\&&-I(W;Y_{2li}\mid Y_l^{i-1}Y_{[1,l-1]}^n Y_{2l[i+1]}^n Y_{2[l+1,L]}^n)+\epsilon_{n}\notag \\
&=& \frac{1}{n}\sum_{l=1}^L \sum_{i=1}^n I(W Y_{2l[i+1]}^n Y_{2[l+1,L]}^n ;Y_{li}\mid \notag\\&&  Y_l^{i-1}Y_{[1,l-1]}^n Y_{2l[i+1]}^n Y_{2[l+1,L]}^n) \notag \\&&-I(WY_l^{i-1}Y_{[1,l-1]}^n Y_{2l[i+1]}^n Y_{2[l+1,L]}^n;Y_{2li}\mid\notag \\&& Y_l^{i-1}Y_{[1,l-1]}^n Y_{2l[i+1]}^n Y_{2[l+1,L]}^n)+\epsilon_{n}\notag \\
&\le& \frac{1}{n} \sum_{l=1}^L \sum_{i=1}^n I(W Y_{2l[i+1]}^n Y_{2[l+1,L]}^n ;Y_{li}Y_{1li}\mid \notag\\&& Y_l^{i-1}Y_{[1,l-1]}^n Y_{2l[i+1]}^n Y_{2[l+1,L]}^n) \notag \\&&-I(WY_l^{i-1}Y_{[1,l-1]}^n Y_{2l[i+1]}^n Y_{2[l+1,L]}^n;Y_{2li}\mid\notag \\&& Y_l^{i-1}Y_{[1,l-1]}^n Y_{2l[i+1]}^n Y_{2[l+1,L]}^n)+\epsilon_{n}\notag \\
\label{u21}
&\overset{(f)}{=}&  \frac{1}{n} \sum_{l=1}^L \sum_{i=1}^n I(V_{1li};Y_{li}Y_{1li}\mid U_{li} V_{2li} )\notag\\&&-I(V_{1li} V_{2li} ;Y_{2li}\mid U_{li})+\epsilon_{n} \\
\label{u2}
&\overset{(g)}{\le}& \sum_{l=1}^L I(V_{1l};Y_l Y_{1l}\mid U_{l} V_{2l} )-I(V_{1l}V_{2l} ;Y_{2l}\mid U_{l})\notag\\&&+\epsilon_{n}
\end{eqnarray}
where $(f)$ and $(g)$ follow from the above definition.

\vspace{1em}
\textbf{b)} We now bound the rate $R$ as follows.
\begin{eqnarray}
\label{achrate}
nR&=& {H}(W)\notag \\
&=& I(W;Y_{[1,L]}^n)+H(W \mid Y_{[1,L]}^n)\notag\\
&\overset{(h)}{\le}&  I(W;Y_{[1,L]}^n)+ n \epsilon_{n}\notag \\
&=& \sum_{l=1}^L I(W;Y_{l}^n \mid Y_{[1,l-1]}^n)+n \epsilon_{n}\notag\\
&=&  \sum_{l=1}^L\sum_{i=1}^n I(W;Y_{li}\mid Y_l^{i-1} Y_{[1,l-1]}^n)+ n \epsilon_{n} \notag\\
&=&  \sum_{l=1}^L\sum_{i=1}^n H(Y_{li}\mid Y_l^{i-1} Y_{[1,l-1]}^n)\notag\\&&-H(Y_{li}\mid W Y_l^{i-1} Y_{[1,l-1]}^n)+ n \epsilon_{n} \notag\\
&\overset{(i)}{\le}&  \sum_{l=1}^L\sum_{i=1}^n H(Y_{li})-H(Y_{li}\mid W Y_l^{i-1} Y_{[1,l-1]}^n)+ n \epsilon_{n} \notag\\
&\overset{(j)}{\le}&  \sum_{l=1}^L\sum_{i=1}^n H(Y_{li})\notag\\&&-H(Y_{li}\mid W Y_l^{i-1} Y_{[1,l-1]}^n Y_{2l[i+1]}^n Y_{2[l+1,L]}^n)+ n \epsilon_{n} \notag\\
&=&  \sum_{l=1}^L\sum_{i=1}^n I(W Y_l^{i-1} Y_{[1,l-1]}^n Y_{2l[i+1]}^n Y_{2[l+1,L]}^n;Y_{li}) + n \epsilon_{n} \notag\\
&=&  \sum_{l=1}^L\sum_{i=1}^n I(V_{1li},V_{2li} ;Y_{li})+ n \epsilon_{n}.
\end{eqnarray}
Hence, we have
\begin{eqnarray}
\label{u3}
R&\le& \frac{1}{n} \sum_{l=1}^L\sum_{i=1}^n I(V_{1li},V_{2li} ;Y_{li})+ \epsilon_{n} \notag\\
 &\le& \sum_{l=1}^L I(V_{1l},V_{2l} ;Y_{l})+  \epsilon_{n}
\end{eqnarray}
where $(h)$ follows from Fano's inequality; $(i)$ and $(j)$ follows from the fact that conditioning reduces entropy.

We can also bound the rate $R$ as follows
\begin{eqnarray}
\label{achrate2}
nR&=& {H}(W)\notag \\
&=& I(W;Y_{[1,L]}^n)+H(W \mid Y_{[1,L]}^n)\notag\\
&\overset{(k)}{\le}&  I(W;Y_{[1,L]}^n)+ n \epsilon_{n}\notag \\
&=& \sum_{l=1}^L I(W;Y_l^n\mid Y_{[1,L]}^n)+H(W \mid n \epsilon_{n})\notag\\
&=&    \sum_{l=1}^L\sum_{i=1}^n I(W;Y_{li}\mid Y_l^{i-1} Y_{[1,l-1]}^n)+ n \epsilon_{n} \notag\\
&\le&  \sum_{l=1}^L\sum_{i=1}^n I(W;Y_{1li}Y_{li}\mid Y_l^{i-1} Y_{[1,l-1]}^n)+ n \epsilon_{n} \notag\\
&=&  \sum_{l=1}^L\sum_{i=1}^n H(Y_{1li}Y_{li}\mid Y_l^{i-1} Y_{[1,l-1]}^n)\notag \\&&-H(Y_{1li}Y_{li}\mid W Y_l^{i-1} Y_{[1,l-1]}^n)+ n \epsilon_{n} \notag\\
&\overset{(l)}{\le}&  \sum_{l=1}^L\sum_{i=1}^n H(Y_{1li}Y_{li}\mid Y_l^{i-1} Y_{[1,l-1]}^n)\notag \\&&-H(Y_{1li}Y_{li}\mid W Y_l^{i-1} Y_{[1,l-1]}^n Y_{2l[i+1]}^n Y_{2[l+1,L]}^n)\notag\\&&+ n \epsilon_{n} \notag\\
&=&  \sum_{l=1}^L\sum_{i=1}^n I(W Y_{2l[i+1]}^n Y_{2[l+1,L]}^n ;Y_{1li}Y_{li} \mid Y_l^{i-1} Y_{[1,l-1]}^n ) \notag\\&&+ n \epsilon_{n} \notag\\
&=&  \sum_{l=1}^L\sum_{i=1}^n I(V_{1li};Y_{1li}Y_{li}\mid V_{2li})+ n \epsilon_{n}.
\end{eqnarray}
Hence, we have
\begin{eqnarray}
\label{u4}
R &\le& \frac{1}{n}\sum_{l=1}^L\sum_{i=1}^n I(V_{1li};Y_{1li}Y_{li}\mid V_{2li}) +  \epsilon_{n} \notag\\
  &\le& \sum_{l=1}^L I(V_{1l} ;Y_{l}Y_{1l}\mid V_{2l})+  \epsilon_{n}
\end{eqnarray}
where $(k)$ follows from Fano's inequality; and $(l)$ follows from the fact that conditioning reduces the entropy.
 
Therefore an outer bound on the achievable rate equivocation region is given by the following set:
\begin{eqnarray}
\bigcup \bigg\{(R,R_e)\:  \text{that satisfy}\:  \eqref{u1},\eqref{u2},\eqref{u3},\eqref{u4}\bigg\}
 \end{eqnarray}
where the union is over all probability distributions $p(u_{[1,L]},v_{1[1,L]},v_{2[1,L]},x_{1[1,L]},x_{2[1,L]},y_{[1,L]},y_{1[1,L]},y_{2[1,L]})$. Finally we note that the terms in  \eqref{u1}, \eqref{u2}, \eqref{u3}, and \eqref{u4} depend on the probability distribution $p(u_{[1,L]},v_{1[1,L]},v_{2[1,L]},x_{1[1,L]},x_{2[1,L]},y_{[1,L]},y_{1[1,L]},y_{2[1,L]})$ only through $p(u_l,v_{1l},v_{2l},x_{1l},x_{2l},y_{l},y_{1l},y_{2l})$. Hence, there is no loss of optimality to consider only those distributions that have the form
\begin{eqnarray}
\label{mark}
\prod_{l=1}^L \bigg[p(u_l,v_{1l},v_{2l})p(x_{1l},x_{2l}\mid u_l,v_{1l},v_{2l}) \notag\\.p(y_{l},y_{1l},y_{2l}\mid x_{1l},x_{2l})\bigg].
\end{eqnarray}
 This completes the proof of Theorem 1.
 
\setlength{\arraycolsep}{5pt}
 \section*{Acknowledgments}
This work is supported by the European Commission in the framework
of the FP7 Network of Excellence in Wireless Communications. The authors would
also like to thanks the concerted Action SCOOP for funding.
\bibliographystyle{IEEEtran}
\bibliography{IEEEsecrecy}

\begin{thebibliography}{1}
\providecommand{\url}[1]{#1}
\csname url@samestyle\endcsname
\providecommand{\newblock}{\relax}
\providecommand{\bibinfo}[2]{#2}
\providecommand{\BIBentrySTDinterwordspacing}{\spaceskip=0pt\relax}
\providecommand{\BIBentryALTinterwordstretchfactor}{4}
\providecommand{\BIBentryALTinterwordspacing}{\spaceskip=\fontdimen2\font plus
\BIBentryALTinterwordstretchfactor\fontdimen3\font minus
  \fontdimen4\font\relax}
\providecommand{\BIBforeignlanguage}[2]{{%
\expandafter\ifx\csname l@#1\endcsname\relax
\typeout{** WARNING: IEEEtran.bst: No hyphenation pattern has been}%
\typeout{** loaded for the language `#1'. Using the pattern for}%
\typeout{** the default language instead.}%
\else
\language=\csname l@#1\endcsname
\fi
#2}}
\providecommand{\BIBdecl}{\relax}
\BIBdecl

\bibitem{11}
T.~Cover and A.~E. Gamal, ``Capacity theorems for the relay channel,''
  \emph{IEEE Transactions on Information Theory}, vol.~25, no.~5, pp. 572--584,
  Sep. 1979.

\bibitem{12}
J.~N. Laneman, D.~N.~C. Tse, and G.~W. Wornell, ``Cooperative diversity in
  wireless networks: Efficient protocols and outage behavior,'' \emph{IEEE
  Transactions on Information Theory}, vol.~50, no.~12, pp. 3062--3080, Dec.
  2004.

\bibitem{10}
L.~Lai and H.~E. Gamal, ``The relay–eavesdropper channel: {C}ooperation for
  secrecy,'' \emph{IEEE Transactions on Information Theory}, vol.~54, no.~9,
  pp. 4005--4019, Sep. 2008.

\bibitem{15}
Y.~Liang, H.~V. Poor, and S.~{Shamai (Shitz)}, ``Secure communication over
  fading channels,'' \emph{IEEE Transactions on Information Theory}, vol.~54,
  no.~6, pp. 2470--2492, Jun. 2008.

\bibitem{19}
A.~Khisti and G.~W. Wornell, ``Secure transmission with multiple antennas {II}:
  The {MIMOME} wiretap channel,'' \emph{to appear, IEEE Transactions on
  Information Theory}.

\bibitem{21}
F.~Oggier and B.~Hassibi, ``The secrecy capacity of the {MIMO} wiretap
  channel,'' \emph{Available: http://arxiv.org/abs/0710.1920}.

\bibitem{16}
V.~Aggarwal, L.~Sankar, A.~R. Calderbank, and H.~V. Poor, ``Secrecy capacity of
  a class of orthogonal relay eavesdropper channels,'' \emph{EURASIP Journal on
  Wireless Communications and Networking, Special Issue on Wireless Physical
  Layer Security}, vol. 2009.

\bibitem{22}
R.~G. Gallager, \emph{Information theory and reliable communication}.\hskip 1em
  plus 0.5em minus 0.4em\relax New York:Wiley, 1968.

\bibitem{9}
I.~Csisz\'{a}r and J.~K{\"{o}}rner, ``Broadcast channels with confidential
  messages,'' \emph{IEEE Transactions on Information Theory}, vol.~24, no.~3,
  pp. 339--348, May. 1978.

\end{thebibliography}
\end{document}